# In Elections, Irrelevant Alternatives Provide Relevant Data

Richard B. Darlington
Cornell University

## Abstract

The electoral criterion of independence of irrelevant alternatives (IIA) states that a voting system is unacceptable if it would choose a different winner if votes were recounted after one of the losers had dropped out. But IIA confuses the irrelevant alternative (the candidate who withdrew) with the data which was generated by that candidate's presence in the race. This paper reports a wide variety of simulation studies which consistently show that data from dropout candidates can be very useful in choosing the best of the remaining candidates. These studies use well-validated spatial models in which the most centrist candidates are considered to be the best candidates. Thus IIA should be abandoned. The majority judgment (MJ) voting system was created specifically to satisfy IIA. Some of these studies also show the substantial inferiority of MJ to other voting systems. Discussions of IIA have usually treated dropouts as strictly hypothetical, but our conclusions about the usefulness of dropout data may apply even to real dropouts.

# Introduction

A voting system violates the electoral criterion of "independence of irrelevant alternatives" (IIA) if an election's winner may change when votes are recounted after one of the losers drops out. IIA was treated prominently by Arrow (1951), dismissed by Tideman (2006, p. 132) and others, and defended anew by Balinski and Laraki (2010), who considered it to be a major advantage of their majority judgment (MJ) electoral system. Under the name "subset choice condition" (SCC), a major review of voting systems and criteria (Felsenthal, 2012, p. 23) listed IIA as one of 15 electoral criteria which at least needs to be considered seriously, and that's what we are doing. Most of this paper reports computer simulations questioning the usefulness of IIA and MJ, but we start with several other points.

The central problem with IIA is that it confuses the "irrelevant alternative" (the candidate who withdrew) with the data collected because that candidate was in the race. Any candidate has two roles: as a potential winner, and as a "foil" or standard of comparison for other candidates. If a candidate withdraws after votes were cast, they're no longer a potential winner, but they may still be useful as a foil. To see why, imagine a sports league with three teams in which each pair of teams plays 9 games. Team A has won all 9 of its games against B, and B has won all its games against C, but C has beaten A 5 games to 4. Thus teams A, B, and C have won 13, 9, and 5 games respectively. From these results we must name a league champion. If we choose A, that would make B one of the losers. But if B were suspended for hazing and became ineligible, and we therefore ignored the results of B's games, we would be forced to choose C, who had beaten A 5 games to 4. Thus the initial choice of A violates IIA. But is that reasonable? Team B's ineligibility doesn't mean that the results of its 18 games are irrelevant to the choice between A and C. In fact, choosing A violates IIA even if B hasn't withdrawn, because we can see what would happen if B *did* withdraw.

Sports are not elections, so we focus now on elections. MJ plays a major role in our computer simulations, so we describe it here. In MJ, voters rate candidates on a scale with verbal labels like "excellent," "acceptable," and "poor." The candidate with the highest median rating is the winner. If there is a tie, the ratings for each tied candidate are sorted from high to low. These sorted columns (one for each of the tied candidates) can be placed side by side, forming a matrix. We find the tie-free row nearest the median row of that matrix, and the candidate with the highest rating in that row is the winner. If two such rows are equidistant from the median row, the one below the median row is used. This system will break all ties unless two candidates have absolutely identical distributions of ratings. Since the relative size of two tie-adjusted medians does not change when a third candidate drops out, MJ satisfies IIA. As in many other systems, if a voter fails to rate a candidate, the voter is presumed to have placed that candidate at the very bottom of the scale, since the voter didn't even care enough about the candidate to form a clear opinion about him or her.

The many puzzling paradoxes and conflicts in electoral theory are mostly absent from two-candidate races, but MJ can conflict with simple majority rule (MR) even in those races. For instance, suppose that 5 voters rate candidate A on a 4-point scale as 1 2 3 4 4, where 4 is high. With voters in the same order, their ratings of B are 2 3 4 1 1. Then A's median rating is 3 whereas B's is 2, so A wins by MJ. But the first 3 voters prefer B to A whereas only the last 2 prefer A to B, so B wins by MR.

A Condorcet winner is a candidate who wins all their two-way races by majority rule. A Condorcet paradox occurs if there is no Condorcet winner. MR conflicts with IIA when there is a Condorcet paradox. To see why, suppose a voting system picks some winner in a 3-candidate race with a paradox, and the candidate who lost to that winner then drops out, leaving only the winner and the one who had beaten him or her. The latter candidate would be chosen by MR, so any multi-candidate voting system must violate IIA if the system reduces to MR in two-candidate elections, much as in our sports example. The only major voting systems which don't reduce to MR are MJ and range voting (rangevoting.org), so all the others violate IIA.

It might be suggested that any voting system could be made to satisfy IIA simply by specifying that votes *not* be recounted if a loser drops out. A response to that is that the whole purpose of IIA is not to say how to handle real dropouts, but to assess voting systems on the assumption they do not include any such extra clause. This paper's final section explores that issue further.

After studying hundreds of real-world elections, Tideman and Plassmann (2012, p. 245) concluded that spatial models fit real-world data far better than any other known model. This paper reports two sets of computer simulations, all using two-dimensional spatial models. Each dimension in these models represents an opinion dimension such as liberalism versus conservatism in domestic or foreign policy. Each candidate and each voter is represented by a dot in this space. Each voter is presumed to rank the candidates by their distance from the voter, with the closest candidates ranked highest. The "best" candidates are presumed to be the ones nearest the center (mean) of all the voters. Even if spatial models like these were to ultimately turn out to be unrealistic, recall that IIA and other electoral criteria are intended to supersede questions of reality, and apply to any possible world. Spatial models certainly represent a possible world, so IIA should be dismissed as a criterion if studies using the models contradict it.

Many studies of this sort draw voters randomly from a univariate, bivariate, or multivariate normal distribution, drawing new voters for each trial, but leave the candidates fixed from trial to trial. The present studies draw both voters and candidates anew on each trial, drawing both from a standardized bivariate normal distribution with independence between the two dimensions. This is done to study a wide variety of possible configurations of candidates as well as voters. Spatial positions, centrism values, and closeness values were always computed to 16 decimal digits, so no two candidates were ever exactly tied on centrism or on closeness to any voter.

## Twelve Studies Using Data from Losers

### Procedures

Let $c$ denote the number of candidates in a particular study; different studies reported here used $c$-values of 5, 10, and 20. Each trial had 100 voters. In each trial, MJ (with its very effective tiebreaker) was used to select two "finalists." The first finalist was the MJ winner; the second was the MJ winner after the first finalist had been removed. Then a two-way majority-rule race was run between each of the finalists and each of the remaining $c$-2 candidates, making 2($c$-2) such races in all. A trial was kept for further analysis only if both finalists beat all their opponents in these races. Thus in all trials kept for later analysis, the finalists were the top two candidates by both MJ and MR. The intent here was to use only trials in which it was crystal clear that the two finalists fully deserved to be in that position and the remaining candidates could reasonably be called "losers." Each study used 1 million trials which met these criteria.

4The studies were of four types. Within each type there were three studies, using 5, 10, and 20 candidates respectively, thus making 12 studies altogether. This was done to see how much the study results depended on the details of study design. In Type 1 the closeness values themselves were used as voter ratings of candidates which were entered into electoral systems to choose winners. In Type 2 the closeness values were rounded to the nearest integer before being entered into the electoral systems. This simulates the type of ballot used in MJ. These ratings ranged from 1 to 9, although over 99% of all ratings were from just 5 to 9. This rounding produced many ties between ratings, either across candidates within a voter, or within candidates across voters, whereas no two ratings were ever tied in Type 1. In studies of Type 3 the rounding was omitted, but mutually independent standard normal random-error values were added to the closeness values to create the ratings used by the electoral systems. This simulated the fact that different voters may perceive the same candidate differently, either because of deliberate deception by candidates or inattention by voters. In studies of Type 4 the ratings of Type 3 were rounded to the nearest integer to create the final ratings. Thus the four types of study are arranged in a 2 x 2 matrix: with or without rounding, and with or without extra random error added.

In each trial, four electoral systems (two unique to this study) were used to choose a final winner from the two finalists. The first system was MJ itself. This required no new analysis; the MJ winner was the first finalist selected. The second method was simple majority rule (MR) between the two finalists. The third method, called QB for quasi-Borda, consisted of computing each finalist's mean margin of victory in two-way races against the losers, and choosing the finalist with the higher mean. Recall that we're using only trials in which the two finalists each beat all other candidates in two-way races. In the fourth electoral system, the winner was the finalist whose smallest margin of victory against the losers exceeded the other finalist's smallest margin. I called this system quasi-minimax or QM. In the well-known minimax system, a candidate wins if he or she beats all other candidates by MR in two-way races. If no such winner exists, the winner is the one whose largest loss is smallest. Thus QM is as similar to minimax as a system using only "irrelevant" data can be.

Thus we have two methods (MJ and MR) which use only data which IIA considers "relevant," and two others (QB and QM) using only data which IIA considers "irrelevant" because that data was available only because the losers had been in the race. IIA considers that data so "poisonous" that it labels any voting system unacceptable if it uses even a bit of that data. But QB and QM use *only* that data. A method was considered to have "hit" on any particular trial if it chose the more centrist of the two finalists, and to have missed if it chose the other.

With its tiebreaker, MJ occasionally produces ties with very few voters, but essentially never produces ties in samples of 100 voters each. But the other three methods all produce noticeable numbers of ties in samples this size. Ties are rare in all these methods when the number of voters is larger. Thus ties can be considered an artifact produced by the need to use small samples to efficiently generate millions of trials. To simulate the absence of this artifact, I discarded any trial in which any of the four electoral systems generated a tie. In each of the 12 studies, the computer ran until 1 million tie-free trials had been found.

**Data Analysis**

Results appear in Table 1, which shows 10 x 12 or 120 percentages. In each set of trials I computed two types of percentage. One was the percentage of times a particular electoral system selected the more centrist of the two finalists. The other type of percentage was used to compare two voting systems. For such comparisons I ignored all trials in which the two systems picked the same winner, and reported just the percentage of the remaining trials in which one of them (the system named first in the relevant line of Table 1) beat the other system by picking the more centrist candidate. The naming order of the two systems was chosen so that throughout the table, values over 50 contradict IIA.

The standard error of any percentage is $100\sqrt{[p(1-p)/n]}$, where $p$ is the corresponding proportion (i.e., percentage/100) and $n$ is the sample size. For entries in lines 1-2, $n$ is 1 million. For any value of $n$, the standard error is maximized when $p = 0.5$. Thus the largest possible standard error for any entry in lines 1-2 is $100 \cdot \sqrt{(0.5 \cdot 0.5/1000000)}$ or 0.05. For entries in lines 4-10, $n$ is the number of trials on which the two methods being compared picked different winners. Thus $n$ is different for each entry in those lines. The value of $n$ also varied in line 3, for reasons explained later. The formula just given was used to estimate the standard error for each of the 96 percentages in lines 3-10. Only three of these standard errors exceeded 0.3, and those were all in lines 8-10, in which MR is compared to QB or QM. Even in those lines, none exceeded 0.8. Because the figures in Table 1 are rounded to the nearest integer (when expressed as percentages rather than proportions), and the standard errors of all but one of these percentages are below 0.5, all those tabular entries can be thought of as essentially free of sampling error. The only exception is the entry of 71 in line 10 and column 3. That value's standard error was 0.76, so that one value could range from 70 to 72. That range is still trivial, so we'll treat all 120 entries in Table 1 as essentially exact.

**Results and Discussion**

Each column in Table 1 gives results for one of the 12 studies described above. IIA and MJ would predict that all 120 percentages in Table 1 would be 50 or below, so the farther any percentage falls above 50, the less consistent it is with IIA or MJ or both.

Before considering individual entries in Table 1, here are a few comments about the table as a whole. The table's entries differ substantially from column to column. This shows that as intended, the 12 studies did study the four electoral systems (MR, MJ, QB, QM) under noticeably different conditions. Because these entries contain essentially no sampling error, even a single value above 50 raises doubts about the value of IIA or MJ or both. But 93 of the 120 entries exceed 50, and range as high as 92.

The 84 entries in lines 1-7 are the ones most relevant to MJ and IIA. Only 6 of those values are below 50; they are underlined. The mean of all values in those lines is 65.7, and the median is 64.5. High entries are scattered all over those 7 lines; in those lines, every line and every column contains at least one entry of 65 or higher.

We now discuss the parts of Table 1 according to the scientific questions they answer.

**Table 1.** Entries are percentages of positive results. In lines 1-3 a positive result occurs when the named system chooses the more centrist of the two finalists. In lines 4-10 the only trials counted were those in which the two systems named picked different winners. In those lines a result was called positive if the winner picked by the first-named system was the more centrist of the two. Throughout the table, values over 50 contradict IIA.

|      |            | No extra error term ||||||  Extra error term ||||||  |
|      |            | Type 1: No rounding ||| Type 2: Rounding ||| Type 3: No rounding ||| Type 4: Rounding ||| Line max |
| Line | $c$ =      | 5  | 10 | 20 | 5  | 10 | 20 | 5  | 10 | 20 | 5  | 10 | 20 |    |
|---|---|---|---|---|---|---|---|---|---|---|---|---|---|---|
| 1 | QB         | 72 | 71 | 69 | 81 | 77 | 70 | 66 | 56 | <u>49</u> | 70 | 61 | 54 | **81** |
| 2 | QM         | 75 | 74 | 70 | 84 | 78 | 68 | 54 | 63 | <u>46</u> | 70 | 59 | 52 | **84** |
| 3 | QBQM       | 86 | 89 | 90 | 92 | 88 | 79 | 80 | 72 | 66 | 80 | 65 | 65 | **92** |
| 4 | QB > MJ    | <u>36</u> | 57 | 71 | <u>45</u> | 62 | 67 | 59 | 63 | 63 | 60 | 63 | 62 | **71** |
| 5 | QM > MJ    | <u>43</u> | 63 | 71 | 55 | 67 | 64 | 52 | 55 | 55 | 55 | 55 | 55 | **71** |
| 6 | QBQM > MJ  | <u>49</u> | 72 | 83 | 62 | 74 | 70 | 61 | 65 | 64 | 63 | 65 | 63 | **83** |
| 7 | MR > MJ    | 80 | 83 | 84 | 71 | 68 | 64 | 57 | 56 | 55 | 59 | 57 | 55 | **84** |
| 8 | QB > MR    | 27 | 44 | 60 | 15 | 23 | 32 | 52 | 58 | 59 | 52 | 57 | 58 | **60** |
| 9 | QM > MR    | 32 | 49 | 59 | 18 | 27 | 32 | 46 | 49 | 50 | 47 | 49 | 50 | **59** |
| 10 | QBQM > MR | 35 | 58 | 71 | 20 | 33 | 44 | 54 | 59 | 60 | 54 | 58 | 59 | **71** |
|   | Column max | 86 | 89 | 90 | 92 | 88 | 79 | 80 | 72 | 66 | 80 | 65 | 65 | **92** |

*Do the two "irrelevant" methods perform better than chance?*

Line 1 in Table 1 shows, for each of the 12 studies, the percentage of trials in which method QB hit – that is, QB selected the more centrist of the two finalists. IIA predicts these values to be around 50. All but one are over 50, 9 are over 60, and one is 81. Comparable results for method QM appear in line 2. In that line, all but one of the percentages are over 50, 8 are over 60, and the highest one is 84.

Since methods QB and QM both use only data which IIA considers irrelevant, IIA must predict that performance will be at chance levels even when the two methods agree. Line 3 shows the hit percentages for QB and QM when we examine only trials in which those two methods pick the same candidate. All 12 of those values are 65 or higher, 5 are 85 or higher, and the highest one is 92.

Thus QB and QM perform far above chance, despite using only data which IIA considers irrelevant.

*Do those "irrelevant" methods actually outperform MJ and MR?*

MJ was designed to be the best electoral system which fully satisfies IIA. Thus a tougher test for QB and QM is whether they outperform MJ. Line 4 shows the percentages of trials on which QB outperformed MJ when the two methods picked different winners. Again, chance is 50. Ten of the 12 values in that line exceed 50, 8 are 60 or higher, and the highest one is 71. Line 5 presents similar results for method QM. In that line, 11 of the 12 values exceed 50, 4 are 60 or higher, and the highest one is 71. When we define method QBQM as in the previous subsection, line 6 shows that it outperforms MJ in 11 of the 12 studies, all of those 11 percentages exceed 60, and the highest one is 83. Thus the "irrelevant" methods outperform MJ with remarkable consistency. The "irrelevant" methods even often outperform MR; 18 of the 36 entries in lines 8-10 exceed 50, rising up to 71.

### *Is MJ at least the best of IIA's two "relevant" methods?*

This question is addressed by line 7. All 12 entries in that line exceed 50, 6 of them exceed 60, and the highest one is 83. Thus MR consistently outperforms MJ.

## Candidates who Actually Withdraw

IIA is usually understood to be a tool for assessing voting systems, not for dealing with candidates who actually withdraw from a race. This section reports some surprising simulation findings about the data from candidates who (in the computer simulations) actually withdraw. Our purpose is not so much to provide actual advice to election managers as to highlight some very surprising results which appear in these simulations. We consider three cases: (1) winners who withdraw when there is no Condorcet paradox, (2) winners who withdraw when there is one, (3) losers who withdraw when there is one. There is no need to discuss the case of losers who withdraw when there is no Condorcet paradox, since in that case there is clearly no purpose in any further analysis. We consider only the case of three candidates. All these studies used spatial models similar to those described earlier. Each candidate's "centrism" was computed as already described, and the most centrist candidates were considered the best ones.

Study 1 in this section generated 100,000 paradox-free trials, all with 3 candidates and 75 voters. In each trial we imagine that the Condorcet winner withdraws after votes had been counted. We consider three strategies for picking a new winner. Strategy HH, for "head to head," examines only the two-way race between the two remaining candidates, and selects the majority-rule winner in that race. Let MM stand for "minimax modified." For each of the two remaining candidates, strategy MM identifies that candidate's largest loss in their two two-way races, and names as winner the candidate for whom this largest loss was smaller. For those same two candidates, strategy WM (for "winner's margin") looks at their margins of loss against the original winner, and names as winner the one with the smaller margin of loss.

For each pair of strategies we consider only the trials in which the two strategies picked different winners. The method picking the more centrist winner is considered to have won that trial. Strategy WM beat HH 94,041 trials to 5959. WM beat MM 93,979 trials to 6021. HH beat MM 13,742 to 2654. Thus WM is overwhelmingly the best strategy by our definition, though I'm not aware that it has ever been proposed for dealing with this situation. HH would of course be the normal course of action. It did actually outperform MM decisively, though the two approaches picked different winners on only 16,396 of the 100,000 trials.

Study 2 examined 10,000 trials with 3 candidates and 75 voters each, where each trial has a Condorcet paradox. In this study, to make sure the three candidates all differ from each other in essential ways, a trial was kept for analysis only if all three candidates had different LL values, where LL is the size of each candidate's largest loss in two-way races. The minimax system was used to pick a winner; it's the candidate whose LL is smallest. As in Study 1, we imagine the winner drops out after votes are counted. The same three strategies were analyzed as in Study 1: HH, MM, and WM. Now MM beat HH, 2636 to 2395. That difference is not overwhelming, but by a binomial test it is statistically significant at the 0.00072 level two-tailed. MM beat WV more decisively, 4179 to 790. HH beat WM, 6574 to 3426. Thus the most obvious strategy (HH) again outperforms one of its competitors and underperforms the other, though the other two methods have changed places, with MM now performing best and WM worst.

The trials in Study 3 were like those in Study 2, but this time the candidate dropping out was the one who lost to the minimax winner. (All these trials have a Condorcet paradox, which means that each of the three candidates has one two-way win and one loss.) Thus the two remaining candidates were the minimax winner and the candidate to whom that person lost. The minimax winner was the more centrist of those two candidates in 7636 trials and the less centrist in the other 2364 trials, so picking the minimax winner rather than the head-to-head winner appears to be the better strategy.

There is no need for a study on what to do when the candidate dropping out is the one who beat the minimax winner, since the winner should obviously not change in that case.

Thus in all three studies in this section, the most obvious strategy (a head-to-head race between the two remaining candidates) was not the best strategy. We will not pursue here the question of what strategy is actually best for other cases or in real life. The main conclusion to be drawn at this time is that the familiar IIA principle is wrong in more ways than anyone had imagined, because of the useful role that dropout candidates can still play as standards of comparison.